\documentclass[twocolumn,amsmath,amssymb,aps,pra,showpacs,tightenlines]{revtex4}
\usepackage{amsmath}
\usepackage{amssymb}
\usepackage{txfonts}
\usepackage{graphicx}
\usepackage{epstopdf}
\usepackage{dcolumn}
\usepackage{bm}
\usepackage[colorlinks=true,citecolor=blue,linkcolor=magenta]{hyperref}
\begin{document}
\title{Magnon-induced high-order sideband generation}
\author{Zeng-Xing Liu}
\author{Bao Wang}
\author{Hao Xiong}\email{haoxiong1217@gmail.com}
\author{Ying Wu}
\affiliation{School of physics, Huazhong University of Science and Technology, Wuhan 430074, China}
\date{\today}

\begin{abstract}
Magnon Kerr nonlinearity plays crucial roles in the study of optomagnonical system and may bring many interesting physical phenomena and important applications.
In this work, we report the investigation of high-order sideband generation induced by magnon Kerr nonlinearity in an optomagnonical system, which is critically lacking in this emerging research field. We uncover that the microwave driving field plays a significant role in manipulating the generation and amplification of the higher-order sidebands, and more importantly, the sideband spacing can be regulated by controlling the beat frequency between the pump laser and the probe laser, which is extremely eventful for the spacing modulation of the sideband frequency comb. Based on the recent experimental progress, our results will deepen our cognition into optomagnonical nonlinearity and may find fundamental applications in optical frequency metrology and optical communications.
\end{abstract}

\maketitle

\begin{figure}[htbp]
\centering
\includegraphics [width=0.95\linewidth] {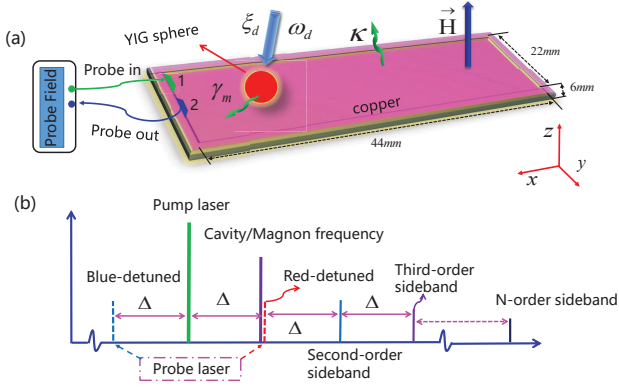}
\caption{(a) Schematic diagram of an optomagnonical system. A millimeter-scale $\rm{YIG}$ sphere driven by a strong microwave source is glued to a three-dimensional copper microwave cavity with dimensions $44\times22\times6$ $\rm{mm^{3}}$ \cite{Bistability}. The system is detected by a weak bichromatic probe laser which consisting of a pump field and a probe field. 
A bias magnetic field with magnetic strength \rm{H} is applied along the z direction. (b) Frequency spectrogram of the optomagnonical system. The frequency of the pump laser is shown by the green line, while the purple line indicates the cavity (magnon) resonance frequency. The blue dashed (blue-detuning) and red dotted lines (red-detuning) represent the probe laser (here we only discuss the latter case), and $\Delta$ is the beat frequency between the pump laser and the probe laser. There are second-, third-, and n-order sidebands generation.}
\label{fig:1}
\end{figure}

Cavity optomagnonics has recently been the focus subject of extensive researches and many interesting phenomena associated with magnon polaritons have been theoretically predicted and experimentally observed \cite{0,1,4,5,7,8,diameter,room0}. A typical optomagnonical system as schematically shown in Fig. \ref{fig:1}(a), in which a millimeter-scale magnetic insulator yttrium iron garnet ($\rm{Y_{3}Fe_{5}O_{12}}$ YIG) \cite{YIG} is loaded in a high-finesse microwave cavity. Under the bias magnetic field, the YIG crystal gives rise to a new polaritons state \cite{polaritons1,polaritons2,magnon} called magnons. The coherent coupling between the magnons and microwave photons \cite{Photons2,Photons3,Photons4,Photons5}, superconducting qubits \cite{superconducting qubit1,superconducting qubit2}, electron \cite{electron1} and phonons \cite{phonon} has been experimentally demonstrated, which can selectively utilize the special advantages from different physical systems to better explore novel phenomena. Recently, a up-and-coming conception magnon-based computing circuits \cite{electron2} has been proposed and the magnons have potential for the implementation of alternative computing concepts \cite{computer}. In addition, cavity-magnon system has opened up avenues of providing a new platform to establish quantum information network thanks to the long coherence time of both magnons and microwave photons \cite{1,4,diameter}. An important example is that the magnon dark modes can be utilized to build a magnon gradient memory \cite{gradient memory} to store information at room temperature, which opened the curtain in the application of quantum information processing.

To our knowledge, however, the current investigation often ignore optomagnonics nonlinear effect which will lead to some fascinating phenomena, such as frequency conversion \cite{soliton} and magnon spintronics \cite{electron2}, so further insight into the nonlinear regime may open up a new and broad prospective for the properties of the optomagnonical system.
Recently, the concept of magnon Kerr nonlinearity, originating from the magnetocrystalline anisotropy in the YIG \cite{anisotropy1,anisotropy2}, has been introduced in a strongly coupled cavity-magnon system, and the nonlinearity-induced frequency shift as well as the bistability of cavity magnon polaritons have been experimentally confirmed \cite{Kerr effect,Bistability}.
These pioneering works establish solid experimental foundation to explore nonlinear characteristics in this emerging field.
As an essential nonlinear phenomenon: sideband generation, however, is still unexplored in optomagnonical system.
The studies of sideband generation, therefore, could deepen our understanding of optomagnonical nonlinearity
and substantially promote the development of this emerging field. Additionally, achieving the feasibility of sideband modulation has extremely important potential applications.

In this work, we investigate high-order sideband generation induced by magnon Kerr nonlinearity in optomagnonical system.
The high-order sideband is a comb-like spectral structure containing a series of equidistant frequency components. As shown in Fig. \ref{fig:1}(b), a pump laser with frequency $\omega_{\rm{l}}$ and a probe laser with frequency $\omega_{p}$ are incident on the cavity field, and then the output field will generate a series of sidebands with equidistant frequency $\mathcal{F}_{n}=\omega_{\rm{l}}\pm n\Delta$ where $\Delta$ is the beat frequency of the pump laser and the probe laser, and n is an integer that represents the order of the sidebands.
Compared with high-order sideband generation in optomechanical system \cite{sideband1,sideband2}, we found that the magnon-induced high-order sideband generation (MHSG) proposed here has some special features. For example, the generation and amplification of MHSG does not satisfy the linear relationship with the power of the control field because the magnon Kerr nonlinearity will reach saturation when the driving power exceeds a certain threshold. An optimal power value, about 12 \rm{mW}, is proposed, which may have potential applications in low-power optical frequency combs. More interestingly, the sideband spacing of MHSG maybe arbitrarily manipulated by harnessing the beat frequency, which is of great significance for improving the measurement accuracy of the frequency-combs-based senor. So our scheme may not only deepen our understanding of optomagnonical nonlinearity but also find applications in optical frequency metrology \cite{metrology1,metrology2,Solit,Solit1,metrology3}.

The physical setup under investigation as schematically shown in Fig. \ref{fig:1}(a). The YIG sphere is directly driven by a microwave source at frequency $\omega_{d}$ and the driving strength $\xi_{d}$. The cavity field is detected by a weak bichromatic probe field and the transmission spectroscopy can be measured by the homodyne technique. Under the rotating wave approximation, the Hamiltonian of the whole system in a frame rotating at frequency $\omega_{\rm{l}}$ of the pump field can be written as \cite{Kerr effect}
\begin{eqnarray}
  \hat{H} &=& \hbar\Delta_{a}\hat{a}^{\dagger}\hat{a}+\hbar\Delta_{b}\hat{b}^{\dagger}\hat{b}
  +\hbar\mathcal{K}\hat{b}^{\dagger}\hat{b}\hat{b}^{\dagger}\hat{b}\nonumber\\
  &&+\hbar\mathcal{G}(\hat{a}\hat{b}^{\dagger}+\hat{a}^{\dagger}\hat{b})
  +i\hbar\xi_{d}(\hat{b}^{\dagger}e^{-i\Delta_{b}t}-\hat{b}e^{i\Delta_{b}t})\nonumber\\
  &&+i\hbar\xi_{\rm{l}}(\hat{a}^{\dagger}-\hat{a})
  +i\hbar\xi_{p}(\hat{a}^{\dagger}e^{-i\Delta t}-\hat{a}e^{i\Delta t}),
\end{eqnarray}
where $\Delta_{a(b)}=\omega_{a(b)}-\omega_{\rm{l}}$ being the frequency detuning of
the cavity (magnon) mode relative to the pump field, and $\Delta$ = $\omega_{p}-\omega_{\rm{l}}$ being the beat frequency between the pump laser and the probe laser. $\hat{a}$ ($\hat{a}^{\dagger}$) and $\hat{b}$ ($\hat{b}^{\dagger}$), respectively, are the annihilation (creation) operators of the microwave cavity mode at frequency $\omega_{a}$ and the magnon mode at frequency $\omega_{b}$. The term $\hbar\mathcal{K}\hat{b}^{\dagger}\hat{b}\hat{b}^{\dagger}\hat{b}$ represents the magnon Kerr effect, where $\mathcal{K}=\mu_{0}\mathcal{K}_{0}\varrho^{2}/(\mathcal{M}^{2}\mathcal{V}_{m})$ with $\mu_{0}$ the magnetic permeability of free space, $\mathcal{K}_{0}$ the first-order anisotropy constant,
$\varrho$ the gyromagnetic ratio, $\mathcal{M}$ the saturation magnetization, and $\mathcal{V}_{m}$ the volume of the YIG sphere \cite{Kerr effect}.
$\mathcal{G}$ is the magnon-photon coupling strength and, here, $\mathcal{G}$ $>$ $\kappa,\gamma_{m}$, i.e., the system falls in the strong-coupling regime.
$\xi_{\rm{l}}$ and $\xi_{\rm{p}}$ are the amplitude of the probe fields.
In this work, we are interested in the mean response of the system, so the operators can be reduced to their expectation values, viz., $\alpha(t)\equiv\langle\hat{a}(t)\rangle$ and $\beta(t)\equiv\langle\hat{b}(t)\rangle$.
\begin{figure}[htbp]
\centering
\includegraphics [width=0.95\linewidth] {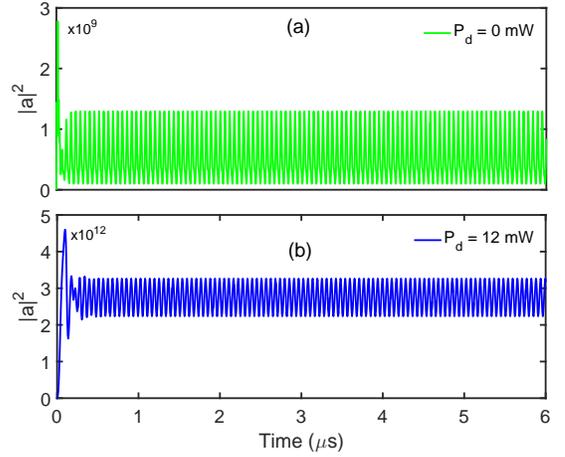}
\caption{Time evolution of photon number $|\alpha|^{2}$ of the microwave cavity field (a) in the absence of microwave driving and (b) in the presence of microwave driving, respectively. The parameters used in numerical simulation are $\omega_{\rm{a}}/2\pi$ = $\omega_{\rm{b}}/2\pi$ = 10.1 $\rm{GHz}$, $\kappa/2\pi$ = 3.8 $\rm{MHz}$, $\gamma_{m}/2\pi$ = 17.5 $\rm{MHz}$, $\mathcal{G}/2\pi$ = 41 $\rm{MHz}$, $\mathcal{K}/\kappa$ = $10^{-10}$, $\Delta_{a}$ = $\Delta_{b}$ = $\Delta$ = $\gamma_{m}$, the power of the bichromatic probe fields are $P_{\rm{l}} = P_{p} = 6.9$ ${\mu W}$. All the parameters are chosen based on the latest experiment \cite{Bistability}.}
\label{fig:0}
\end{figure}
With cavity field and magnon mode damping processes included, the evolution of the optomagnonical system can be described by the Heisenberg-Langevin equations as
\begin{eqnarray}\label{equ:0}
\dot{{\alpha}}&=&(-i\Delta_{a}-\frac{\kappa}{2})\alpha-i\mathcal{G}\beta+\xi_{\rm{l}}+\xi_{p}e^{-i\Delta t},\nonumber\\
\dot{{\beta}}&=&(-i\Delta_{b}-\frac{\gamma_{m}}{2})\beta-i(2\mathcal{K}\beta^{\ast}\beta+\mathcal{K})\beta
  -i\mathcal{G}\alpha+\xi_{d}e^{-i\Delta_{b}t}.
\end{eqnarray}
Here, the mean-field approximation by factorizing averages is used, i.e., $\langle \alpha\beta\rangle\equiv\langle \alpha\rangle\langle \beta\rangle$,
and the quantum noise terms can be dropped safely because their expectation values are zero in the semiclassical approximation.


The photon number $|\alpha|^{2}$ of the microwave cavity field can be obtain by numerically solving Eqs. (\ref{equ:0}), and we plot the time evolution of $|\alpha|^{2}$ in Fig. \ref{fig:0}. We can see that when the system reaches a stable oscillation after a transient process,
the photon number of the microwaves cavity field is relatively small without driving the YIG sphere (Fig. \ref{fig:0} (a)). When the YIG sphere is driven by a strong microwave source, however, the photon number is greatly increased (Fig. \ref{fig:0} (b)).
Considerable magnons will generate from the YIG sphere and the magnon Kerr effect will be exceedingly enhanced by directly driving the YIG sphere \cite{Kerr effect,Bistability}.
Next, we discuss high-order sideband generation of the optomagnonical system. The output field $s_{out}(t)$ can be obtained by using the input-output relation, i.e., $s_{out}(t) = s_{in}(t)-\sqrt{2\kappa}\alpha(t)$, where $s_{in}(t)=\xi_{\rm{l}}e^{-i\omega_{\rm{l}}t}+\xi_{p}e^{-i\omega_{p}t}$
is the bichromatic input laser. Thence, the output field in time domain can be formally expressed as $s_{out}(t)=\sum_{\jmath=o}^{n}\mathcal{A}_{\pm\jmath}e^{-i(\omega_{l}\pm\jmath\Delta)t}$ ($\jmath=0, 1, 2...$), where $\mathcal{A}_{\jmath}$ is the $\jmath$-th transmission coefficient of the output field.
In frequency domain, the output spectrum $\mathcal{S}(\omega)$ can be acquired by performing the fast Fourier transform of $s_{\rm{out}}(t)$, i.e., $\mathcal{S}(\omega) \propto \big|\int_{-\infty}^{\infty}s_{\rm{out}}(t)e^{-i\omega t}dt\big|$
where $\omega$ is the spectroscopy frequency from the microwave cavity field. After doing the fast Fourier transform, the transmission spectrum will appear the second-, third- and higher-order sidebands.

\begin{figure}[htbp]
\centering
\includegraphics [width=0.95\linewidth] {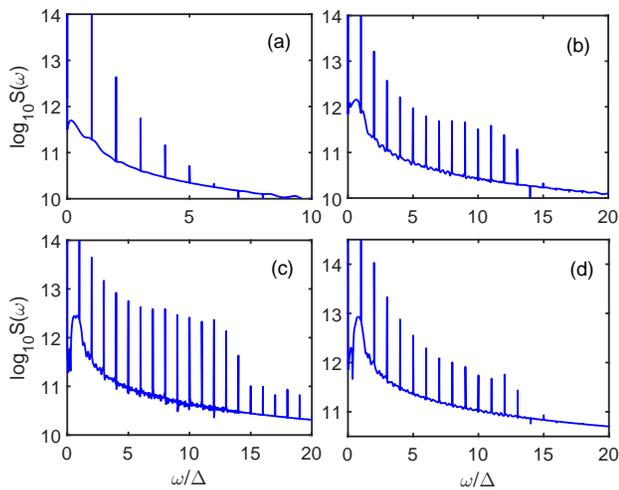}
\caption{The MHSG spectra output from the optomagnonical system are shown with different powers of the microwave driving field $P_{d}$. (a) $P_{d}$ = 1 $\rm{mW}$, (b) $P_{d}$ = 5 $\rm{mW}$, (c) $P_{d}$ = 12 $\rm{mW}$, (d) $P_{d}$ = 19 $\rm{mW}$. The other parameters for simulation are the same as those in Fig. \ref{fig:0}.}
\label{fig:2}
\end{figure}


Figure. \ref{fig:2} shows how the microwave driving field can be used to overmaster the generation and amplification of MHSG. As shown in Fig. \ref{fig:2}(a), the weak microwave driving laser ($P_{d}$ = 1 $\rm{mW}$) only induced a few low-order sidebands. Under this condition, the magnon Kerr-nonlinearity process is very weak and the generated high-order sideband are almost absorbed. In Fig. \ref{fig:2}(b), we increase the microwave driving field power to  $P_{d}$ = 5 $\rm{mW}$, both the order and amplitude of the sidebands have been strengthened, which means that the magnon Kerr nonlinearity reinforces with the increase of microwave driving power. To further increase the power of the microwave driving laser $P_{d}$ = 12 $\rm{mW}$, a series of robust sidebands appear in the frequency spectrum. The spectrum structure of MHSG is that it decreases rapidly for the first few order sidebands and follows by a plateau where all the sidebands have the same strength, and until by a cutoff regime where the amplitude of the sidebands decreases sharply to complete this spectrum. In addition, some nonperturbative signs emerge on the sideband spectrum that is the intensity of the lower-order sidebands have smaller than the higher-order sidebands, which indicates that a strong magnon Kerr nonlinearity be produced from the YIG crystal. Continuously increase the power of the microwave driving field $P_{d}$ = 19 $\rm{mW}$, the MHSG spectrum as shown in Figs. \ref{fig:2}(d), however, both the cutoff order and the amplitude of the higher-order sidebands are reduced rather than enhanced,
which means that the magnon Kerr-nonlinear intensity is no longer increased but suppressed when the microwave driving power exceeds a certain threshold. This result reminds us of the possibility of bridling the magnon nonlinear strength by adjusting the power of the microwave driving field, and further achieve the purpose of harnessing the sideband spectrum may have potential applications in low-power optical frequency combs \cite{combs2,combs3} and optical communications, as \cite{communications} has suggested.

\begin{figure}[htbp]
\centering
\includegraphics [width=0.95\linewidth] {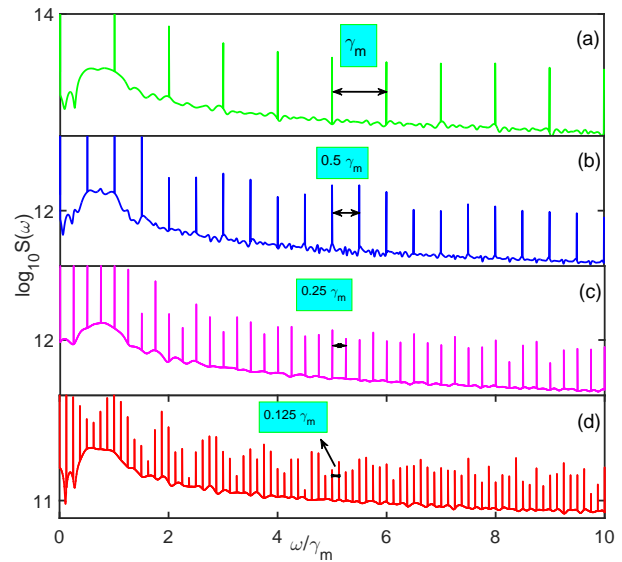}
\caption{The MHSG spectra output from the optomagnonical system are shown with different beat frequency $\Delta$. (a) $\Delta/\gamma_{m}$ = 1, (b) $\Delta/\gamma_{m}$ = 0.5, (c) $\Delta/\gamma_{m}$ = 0.25, (d) $\Delta/\gamma_{m}$ = 0.125. The other parameters for simulation are the same as those in Fig. \ref{fig:0}(b).}
\label{fig:3}
\end{figure}
Realizing sideband-spacing adjustment is of fundamental importance in optical frequency metrology. We uncovered that the sideband spacing in our scheme can be adjusted by regulating the beat frequency between the pump laser and the probe laser.
In Fig. \ref{fig:3}, we plot the MHSG spectra with four different values of the beat frequency $\Delta$. It is clearly shown that as the beat frequency decreases, the spacing of the MHSG spectra becomes narrower and narrower. For convenience, we only discuss the case that the sideband order before $\omega/\gamma_{m} = 10$. For the case of $\Delta/\gamma_{m}$ = 1 in Fig. \ref{fig:3}(a), the number of sidebands is 10 and the spacing of each sideband is $\gamma_m$. For the case of $\Delta/\gamma_{m}$ = 0.5 in Fig. \ref{fig:3}(b), however, the number of sidebands is increased to 20 while the spacing of each sideband is decreased to $0.5\gamma_m$. Further decrease the beat frequency $\Delta_p/\gamma_{m}$ = 0.25 in Fig. \ref{fig:3}(c), the amount of sidebands is increasing fourfold and the spacing of each sideband is as narrow as $0.25\gamma_m$. A more cramped sideband spacing can be obtained as we continue to decrease the beat frequency $\Delta_p/\gamma_{m}$ = 0.125, as Fig. \ref{fig:3}(d) shown.
It is worth pointing out that the bandwidth of the sideband is extremely narrow which can be well explained by the uncertainty relations of time and frequency.
According to the time-frequency uncertainty relation $\Delta\omega\Delta t\sim2\pi$, we can acquire the uncertainty of the frequency as $\Delta\omega\sim2\pi/\Delta t$. Here the applied probe laser is two continuous-wave lasers and it lasts about infinity, i.e., $\Delta t\rightarrow\infty$, therefore, we can obtain the uncertainty of the frequency as $\Delta\omega\rightarrow0$ \cite{sideband2}.
From above discussion, we can see that the beat frequency plays an important role in the modulation of the sidebands spacing, which reminds us of the possibility of implementing the sideband spectrum with arbitrary spacing, so our scheme may has important applications in spectral precision measurement, as \cite{metrology2,metrology0} recommendation.

It is worth emphasizing that the MHSG proposed here has several distinctive advantages. First, our system is a brand new optomagnonical system, which has some unusual merits. For instance, the system has rich magnonic nonlinearities and very low loss for various different information carriers at room temperature \cite{room1}, which may provide environmental conditions for real applications of MHSG. Moreover, the cavity magnonics system possesses the advantages of tunability and compatibility with opto- or electromechanical elements, which provides an excellent platform for quantum state transfer among different physical systems, as \cite{phonon} has shown. Second, we find that the generation and amplification of MHSG can be modulated by the control field and an optimal power region has been proposed, which may find application in low-power frequency combs. In particular, the MHSG spacing is directly determined by the beat frequency between the pump laser and the probe laser, namely, the sideband spectra maybe regulated in arbitrary spacing. So our scheme may trigger substantial advances in practical application in optical frequency metrology and optical communications, as \cite{communications} proposal. Finally, to our knowledge, the investigation of high-order sideband generation induced by magnon Kerr nonlinearity is still unexplored, so this work may deepen our understanding in optomagnonics nonlinearity and may open up a broad prospective for the nonlinear properties of the cavity-magnon system.

Ultimately, it is necessary to evaluate the actual experimental possibilities of the MHSG. The present scheme could be realized under the state-of-the-art experimental conditions. First of all, we note that the Kerr coefficient $\mathcal{K}$ is inversely proportional to $\mathcal{V}_{m}$, thence the Kerr effect will bring strong nonlinearity when using a small YIG sphere. Under the current experimental technology, the YIG crystal can be made into a highly polished sphere with diameter as small as 250 $\mu m$ \cite{phonon}, and such small YIG sphere can bring a sufficiently strong Kerr nonlinearity when we directly driving the YIG sphere. Furthermore, the magnon-photon coupling can be well tuned by adjusting the bias magnetic field \cite{tuned}. In Ref. \cite{Kerr effect}, the external magnetic field is tunable in the range of 0 to 1 T and the frequency of the magnon mode is adjustable ranges from several hundreds of megahertz to 28 GHz, such wide frequency range allows the resonance coupling between the magnons and microwave photons (in Ref. \cite{diameter}, a high cooperativity $\mathcal{C} = 4\mathcal{G}^{2}/\kappa\gamma_{m} = 3.0 \times 10^{3}$ has been achieved). Therefore, our program can easily be implemented under the current experimental conditions.


In summary, an important nonlinear phenomenon: high-order sideband generation induced by magnon Kerr nonlinearity in a strongly coupled optomagnonical system has been explored. We shown that the generation and amplification of the higher-order sidebands can be modulated by bridling the microwave driving field and an optimal power area, about 12 \rm{mW}, is proposed. In addition, we found that the the beat frequency between the pump laser and the probe laser plays critical role in sideband-spacing adjustment. Beyond their fundamental scientific significance, our results may have potential applications in optical frequency metrology and optical communications.


The work was supported by National Natural Science Foundation of China (11774113, 11374116, 11574104, 11375067).

\bigskip
\noindent
\end{document}